\documentclass[twocolumn,floatfix,prb,aps,showpacs]{revtex4}
\usepackage{graphicx,amsmath,amssymb}

\begin{document}

\title{SU(4) composite fermions in graphene: New fractional quantum Hall states}
\author{Csaba T\H oke and J.~K.~Jain}
\affiliation{Department of Physics, 104 Davey Lab, Pennsylvania State University, University Park PA, 16802}
\date{\today}

\begin{abstract}
Theoretical studies of the fractional quantum Hall effect (FQHE) in graphene have so far focused on the plausibility and stability of the previously known FQHE states for the interaction matrix elements appropriate for graphene.  We consider FQHE for SU(4) symmetry, as appropriate for the situation when all four spin and valley Landau bands are degenerate, 
and predict new FQHE states that have no analog in GaAs. These result from 
an essential interplay between the two-fold spin and valley degeneracies 
at fractions of the form  $\nu=n/(2pn\pm 1)$, for $n\geq 3$.  Conditions are outlined for the observation of these states and quantum phase transitions between them; the structure of these states and their excitations is also described.
\end{abstract}

\maketitle

\section{Introduction}

Soon after the first experimental realization of graphene, a single-layer hexagonal form of carbon,
the integral quantum Hall effect (IQHE)\cite{Klitzing80}
was observed by Novoselov \textit{et al.}\ and Zhang \textit{et al.}\cite{Novoselov} at filling factors $\nu_n=4n-2$.
The $\Delta\nu=4$ period is understood straightforwardly 
as a consequence of the four-fold (near) degeneracy of Landau levels (LLs) in graphene; 
the Zeeman energy is small in comparison to the interaction energy scale, and the pseudospin degree of freedom, which represents the two
inequivalent Dirac-cones at the corners of the Brillouin zone, does not couple to external fields if the two sublattices are equivalent.
The offset $-2$ is a consequence of a Dirac-like
effective Hamiltonian, which produces a linear dispersion for low energy 
electronic states \cite{Semenoff}.  Recently, Zhang \textit{et al.}\cite{Zhang} 
have observed quantum Hall plateaus at $\nu=\pm4,\pm1,0$.  Although external 
effects can surely lift the four-fold degeneracy of the low-lying LLs, Yang, Das Sarma
and MacDonald\cite{Yang} have shown that the exchange interaction can
break the SU(4) symmetry spontaneously, and the charged excitations are
skyrmions \cite{Skyrmion} in the $|n|\le3$ Landau levels
at $\tilde\nu=1$ (or, by particle-hole symmetry, at $\tilde\nu=3$) where $\tilde\nu=\nu-\nu_n$ is the filling factor within the Landau level in question.
This has been confirmed \cite{FQHEgraphene} by exact diagonalization for $n\le2$ with $N\le9$ particles.

Although the fractional quantum Hall effect\cite{Tsui82} (FQHE) has not yet been observed in graphene, it has been explored theoretically  
\cite{FQHEgraphene,Apalkov} assuming an SU(2) symmetry, as appropriate, 
for example, when the Zeeman energy is sufficiently high that only 
the two-fold pseudospin degeneracy remains.  In this situation, the graphene 
FQHE problem formally maps 
into the well studied problem of FQHE in GaAs in the zero Zeeman energy 
limit\cite{Wu93,Park1}, with the pseudospin of the former 
playing the role of the spin of the latter.  
In the $n=0$ graphene LL, the interaction pseudopotentials\cite{Haldane} 
are identical to those in GaAs, so the mapping is perfect, and the 
earlier composite-fermion results of Refs.~\onlinecite{Wu93,Park1} 
carry over to graphene with minimal change (spin replaced by pseudospin).
In particular, it follows that FQHE occurs at fractional fillings given by $n/(2pn\pm 1)$, 
with pseudospin singlet states at even $n$ and pseudospin polarized states 
at odd $n$ (fully pseudospin polarized for $n=1$), terminating into a pseudospin 
singlet composite fermion Fermi sea in the limit of $n\rightarrow \infty$ ($\tilde \nu=1/2p$).
The effective interaction in the $n\neq 0$ LLs in graphene interpolates between 
those of the $|n|$th and $(|n|-1)$st LLs for the quadratic dispersion;
the composite fermion formation\cite{Jain89}, and therefore the FQHE, are found to be
almost as strong in the $|n|=1$ LL in graphene as in the lowest LL of GaAs
\cite{FQHEgraphene}.

It is natural to wonder if FQHE with {\em new} structure appears in graphene.  
For this purpose, we explore in this work FQHE including 
the full SU(4) symmetry.
By a combination of exact diagonalization and the CF theory, we find new 
FQHE states which result from an essential interplay between the spin 
and the pseudospin degrees of freedom; such states occur at $\tilde\nu=n/(2n+1)$ 
for $n\geq 3$.   For other states, the energy spectrum of 
the SU(4) problem matches with that of the SU(2) problem, although 
the multiplicities are vastly different.  We show by exact diagonalization that the 
SU(4) symmetry is spontaneously broken at $\tilde\nu=\frac{1}{3}$
just as at $\tilde\nu=1$, i.e., the orbital part of the ground state is antisymmetric and the excitations are skyrmionic.
At $\tilde\nu=\frac{1}{3},\frac{2}{3}$ and $\frac{2}{5}$, the orbital part of ground state is the same as in the SU(2) symmetric system,
but the state is now a highly degenerate SU(4) multiplet.

The outline of the paper is as follows.
We review the methods in Sec.~\ref{methods}.
In Sec.\ \ref{newstates} we generalize the composite fermion theory to SU(4) 
electrons, using Fock's condition and constructing all possible wave functions 
of incompressible, filled $\Lambda$ level states of composite fermions.  We also 
indicate which states have analogs in GaAs and which do not. Section \ref{onethird} is 
concerned with exact diagonalization results for small systems, which are compared to  
the predictions of the CF theory.  Skyrmion-like excitations are also 
discussed.  In Sec.\ \ref{thermo} we obtain thermodynamic energies using the CF theory to determine the thermodynamic ground states.
Section \ref{phasetrans} discusses zero-temperature phase transitions and lists the CF predictions for the parameters where the new FQHE states should be observable.  
The paper is concluded in Section \ref{conclu}.

\section{Model}
\label{methods}

The electronic structure of graphene is well described by a tight binding model\cite{Semenoff},
which gives a vaulted band structure with the valence and conductance 
bands touching in the symmetry protected Dirac (or Fermi) points at the six 
corners of the hexagonal Brillouin zone.
The low energy states, in particular, occupy double cones\cite{Cones} with 
apices at the corners.
Only two of these six double cones are inequivalent, giving rise to a pseudospin degree of freedom.
Denoting spin by $\vec S$ and pseudospin by $\vec P$,
the low energy states are described in the continuum approximation by an effective
Hamiltonian\cite{Semenoff}
\begin{equation}
\label{hamiltonian}
H=v_F
\begin{pmatrix}
\vec\sigma\cdot\vec\Pi & 0\\
0 & (\vec\sigma\cdot\vec\Pi)^{\textrm{T}}\\
\end{pmatrix} + \Delta P_z + g\mu_B \vec B\cdot\vec S,
\end{equation}
that acts on a 4-spinor Hilbert space.
Here $v_F\approx 10^6$ m/s is the Fermi velocity, $\vec\Pi=\vec p+\frac{e}{c}\vec A$,
$\Delta$ is the on-site energy difference between the two sublattices,
$P_i=1\otimes\frac{1}{2}\sigma_i$ is the pseudospin operator, and $\vec S$ is the spin.
Letting $z=x-iy$, measuring distance in units of the magnetic length $\ell_B=\sqrt{\hbar c/eB}$, and using the symmetric gauge $\vec A=(-By/2,Bx/2)$,
this Hamiltonian is diagonalized by the normalized eigenvectors
\begin{eqnarray}
\label{lllpseudo}
\Psi^{(0,m)}_{p=1/2,s}&=&
\begin{pmatrix}
0 \\ \eta_{0,m} \\ 0 \\ 0
\end{pmatrix}\otimes\alpha_s,\\
\Psi^{(0,m)}_{p=-1/2,s}&=&
\begin{pmatrix}
0 \\ 0 \\ \eta_{0,m} \\ 0
\end{pmatrix}\otimes\alpha_s,\\
\label{otherpseudo}
\Psi^{(n\neq0,m)}_{p=1/2,s}&=&\frac{1}{\sqrt 2}
\begin{pmatrix}
-{\rm sgn}(n)i\eta_{|n|-1,m} \\
\eta_{|n|,m} \\ 0 \\ 0
\end{pmatrix}\otimes\alpha_s,\\
\Psi^{(n\neq0,m)}_{p=-1/2,s}&=&\frac{1}{\sqrt 2}
\begin{pmatrix}
0 \\ 0 \\
\eta_{|n|,m} \\
-{\rm sgn}(n)i\eta_{|n|-1,m}
\end{pmatrix}\otimes\alpha_s,
\end{eqnarray}
where $\alpha_{1/2}=\begin{pmatrix}1 \\ 0\end{pmatrix}$, $\alpha_{-1/2}=\begin{pmatrix}0 \\ 1\end{pmatrix}$,
and $\eta_{n,m}$ is the eigenstate of the Hamiltonian with quadratic dispersion in LL $n$ with angular momentum ($L_z$) equal to $m$:
\begin{equation}
\eta_{n,m}(z)=\frac{(-1)^n\sqrt{n!}}{\sqrt{2^{m+1}\pi (m+n)!}}z^m L_n^m\left(\frac{|z|^2}{2}\right) e^{-|z|^2/4}.
\end{equation}
The corresponding energies are
\begin{equation}
\label{landaulevels}
E_{nps}={\rm sgn}(n)\sqrt{\frac{2\hbar v_F^2 eB|n|}{c}} + \Delta p +  g\mu_B Bs.
\end{equation}
In the limit $g\to0,\Delta\to0$, the four vectors
\begin{equation}
\Psi^{(n,m)}_{1/2,1/2},\quad
\Psi^{(n,m)}_{1/2,-1/2},\quad
\Psi^{(n,m)}_{-1/2,1/2},\quad
\Psi^{(n,m)}_{-1/2,-1/2}
\end{equation}
are degenerate, giving rise to an SU(4) internal symmetry.

The SU(4) approximation is valid if $E_Z=g\mu_B B$ and $\Delta$ are small in comparison to the scale of the
interactions, $e^2/(\epsilon l_B)$.
While $\Delta$ is not easy to estimate, it is generally believed to be much smaller than $E_Z$.
One can check that
\[
E_Z\left/\left(\frac{e^2}{\epsilon l_B}\right)\right.=0.001 \epsilon g \sqrt{B[\text{T}]},
\]
where $B[\text{T}]$ is the magnetic field in Tesla.
Obviously, both $\epsilon$ and $g$ depend strongly on the interaction with environment in which the graphene sheet is placed.
As two extremes, one can take $\epsilon_{\text{min}}=(\epsilon_{\text{air}}+\epsilon_{\text{SiO}_2})/2=2.75$ and
$\epsilon_{\text{max}}=15$ (the graphite value).
On the other hand, Zhang \emph{et al.} \cite{Zhang} find $1.7\le g\le2.0$.
With these extreme values, assuming an extremely high field $B=45$ T, we get 
\begin{eqnarray*}
\left(E_Z\left/\left(\frac{e^2}{\epsilon l_B}\right)\right.\right)_{\text{min}}&=&0.032,\\
\left(E_Z\left/\left(\frac{e^2}{\epsilon l_B}\right)\right.\right)_{\text{max}}&=&0.208.
\end{eqnarray*}
Thus the SU(4) symmetric approximation is typically 
valid unless $\epsilon$ happens to be close to the high end of the estimation.

The interaction between eigenstates is conveniently described in terms of pseudopotentials\cite{Haldane}, where the pseudopotential $V_m$ gives the
energy of two electrons in relative angular momentum $m$.
As a consequence of Eqs.~(\ref{lllpseudo}) and (\ref{otherpseudo}), the problem of interacting electrons in the $n$th LL of graphene can be mapped
into a problem of interacting electrons in $\eta_{0,m}$ states with effective pseudopotentials\cite{Apalkov,FQHEgraphene} 
\begin{equation} 
\label{effective}
V_m^{(n)\textrm{gr.}}=\int\frac{d^2k}{(2\pi)^2}\frac{2\pi}{k}F_n(k)e^{-k^2}L_m(k^2),
\end{equation} 
where the form factor $F_n$ is
\begin{equation}
F_0(k)=1,\quad F_n(k)=\frac{1}{4}\left(L_n\left(\frac{k^2}{2}\right) + L_{n-1}\left(\frac{k^2}{2}\right)\right)^2.
\end{equation}

The problem of interacting electrons is conveniently formulated in the
spherical geometry, in which electrons move on the
surface of a sphere and a radial magnetic field is produced by
a magnetic monopole of strength $Q$ at the center.\cite{Haldane,Fano}
Here $2Q\phi_0$ is the magnetic flux through the surface of the sphere; $\phi_0=hc/e$, and
$2Q$ is an integer according to Dirac's quantization condition.\cite{Dirac31}
For quadratic dispersion the single particle states are monopole harmonics\cite{Yang1976} $Y_{Qlm}$, where
$l=Q+n$ is the angular momentum with $n=0,1,\ldots$ being the LL index,
$m=-l,-l+1,\ldots,l$ is the $z$-component of angular momentum.
The Coulomb interaction is evaluated with the chord distance.
The analogous solution for carriers with linear dispersion, described by the massless  Dirac's
equation, is not known for the spherical geometry.  
We proceed to work in the basis of 
the monopole harmonics $Y_{QQm}$ of the lowest LL, while using the 
effective pseudopotentials given in Eq.~(\ref{effective}).
Because we have $F_0(k)=1$, in the $n=0$ Landau level the 
pseudopotentials derived by Fano \textit{et al.}\cite{Fano} can be used.
In other Landau levels, however, we will use
the effective pseudopotentials of the {\em planar} geometry 
(Eq.~(\ref{effective})) on the sphere.
Such an identification is obviously exact in the thermodynamic limit, 
but is usually reasonable also for finite systems.

We neglect the finite thinkness of the two dimensional electron system, which is a much better approximation in graphene than in GaAs heterostructures.
The neglect of Landau level mixing, also made below, is a more tricky assumption because both the LL separation (c.f.\ Eq.~(\ref{landaulevels})) and the interaction energy scale as $\sqrt B$, and are of the same order.  However, their  
relative strengths can be controlled by variation of other parameters (Fermi 
velocity, dielectric constant, etc.), and we will assume in the following that the 
LL separation is large enough to suppress LL mixing.

\section{Composite fermion theory for SU(4) electrons}
\label{newstates}

We will compare the exact diagonalization results to the composite 
fermion theory \cite{Jain89,JainKamilla}, and also use this theory to explore 
the thermodynamic limit.  (Exact diagonalization can be performed  
only for very small systems for SU(4) symmetry.)
The composite fermion (CF) theory maps the strongly
interacting system of electrons in a partially
filled Landau level to a system of weakly interacting particles  
called composite fermions, which are bound states of an electron and an even number 
of quantized vortices. The mapping consists of attaching 
$2p$ quantized vortices of the many-body wave function to each electron.
The quantized vortices produce Berry phases that partly cancel the 
external magnetic field, and consequently composite fermions 
feel a reduced magnetic field $B^\ast=B-2p\rho\phi_0$.
Composite fermions in magnetic field $B^\ast$ fill quantized kinetic energy levels analogous to Landau levels, called $\Lambda$ levels.
If $m$ $\Lambda$ levels are filled, which corresponds 
to a filling factor $\nu=\frac{m}{2pm\pm 1}$, an incompressible quantum liquid 
state results.  The FQHE is thus explained as the IQHE of composite fermions.

According to the CF theory, the wave functions for correlated electrons in the 
lowest Landau level at monopole strength $Q$ are given by 
\begin{equation}
\label{cfwf}
\Psi = P_{\rm LLL}\Phi_1^{2p}\Phi,
\end{equation}
\begin{equation}
\Phi_1 = \prod_{j<k} (u_j v_k - v_j u_k),
\end{equation}
where $\Phi$ is a wave function for $N$ non-interacting electrons at monopole strength 
$q=Q-p(N-1)$, operator $P_{\rm LLL}$ projects a state into the $n=0$ LL,
and $u=\cos(\theta/2)e^{-i\phi /2}$ and $v=\sin(\theta/2)e^{i\phi /2}$.
The procedure taking us from $\Phi$ to $\Psi$ is called ``composite-fermionization,"
which also converts electronic Landau levels into $\Lambda$ levels of composite fermions.
Incompressible ground states are constructed by completely filling the lowest few $\Lambda$ levels in one or more spin bands.
The explicit relation between the monopole strength $Q$, the filling factor $\tilde\nu$ and the particle number $N$ is given below in Eq.~(\ref{mono}).

\subsection{Fock's condition}

The above CF construction is valid for the orbital part of the wave function 
independent of whether the Landau levels are singly, doubly, or quadruply degenerate,
i.e., {whether the low-energy electronic states have no internal (nonorbital) 
symmetry, have SU(2) symmetry, or have SU(4) symmetry, respectively.}
However, while any antisymmetric wave function is valid for a singly degenerate Landau level, 
only those wave functions are legitimate for SU(2) which are eigenstates of $S^2$ and $S_z$.
{For SU($n$) electrons, the wave functions are eigenfunctions of all the Casimir operators and the generators
of the Abelian subalgebra of the Lie algebra of SU($n$).}
Among such functions 
the so-called maximal weight states\cite{Symm} satisfy Fock's 
cyclic condition\cite{Hamermesh}. 
The idea will be to construct valid IQHE wave functions $\Phi$, and then show that 
their ``composite fermionization" produces FQHE wave functions with valid symmetry.

Fock's cyclic condition tells us which orbital wave functions are legitimate
for the highest weight state in the multiplet $[m_1,\dots,m_{n-1}]$ of SU($n$).
We are using the Young tableau notation for the SU($n$) multiplets:
$m_1\ge m_2\ge\dots m_{n-1}\ge 0$ are integers, $m_i$ being the length of the $i$-th row 
of the Young tableau of the representation. (The last $m_i$'s are omitted if zero.)
Recall that Young tableaux are generated from the direct products of the fundamental
representation (a single box) with symmetrization in the rows and antisymmetrization in the columns.

Let $\{\alpha^t\}$ be a basis of the ($n$-dimensional) fundamental representation of SU($n$),
and let $M_t$ be the number of particles assumed to be in the $\alpha^t$ internal state.
Let
\begin{equation}
\Phi'(\{\vec r_j\})={\cal A}\left(
\Phi(\{\vec r_j\})\prod_{t=1}^n\prod_{i=\min_t}^{\max_t}\alpha^t_i
\right),
\end{equation}
where $\min_1=1,\max_1=M_1,\min_2=M_1+1,\max_2=M_1+M_2,\dots$, and $\cal A$ is the antisymmetrizer.
Assume $\Phi(\{\vec r_j\})$ is antisymmetric in each subset $\{\min_t,\dots,\max_t\}$ of its variables.
$\Phi'(\{\vec r_j\})$ is a highest weight state (w.r.t.\ a choice of positive roots) if and only if
it is annihilated by any attempt to antisymmetrize an electron $l$ of type $u$ ($\min_u\le l\le\max_u$)
with respect to the electrons of type $t<u$, i.e.
\begin{equation}
\label{condi}
\left(e-\sum_{k=\min_t}^{\max_t}(k,l)\right)\Phi(\{\vec r_j\})=0,
\end{equation}
where $e$ is the identity and $(k,l)$ permutes indices $k$ and $l$.

Consequently, any orbital wavefunction
\begin{equation}
\Phi=\Phi_1\Phi_2\cdots\Phi_n,
\end{equation}
where $\Phi_s$'s are Slater determinants such that any state $(n,m)$ in $\Phi_s$ is also filled in $\Phi_{s-1}$
(conversely, if $(n,m)$ is empty in $\Phi_s$, then it is also empty in $\Phi_{s+1}$),
is a legitimate highest weight wave function for SU($n$).
This class includes IQHE states ($\Psi_s$ has the lowest $l_s$ Landau level completely filled; $l_s$'s are in decreasing order),
Fermi sea states, and single particle, single hole, and particle-hole pair states above the IQHE and Fermi sea states,
provided that no hole is created in spin state $s$ in a LL that is filled in $\Phi_{s+1}$, and, conversely,
no particle is created in spin state $s$ in a LL that is unfilled in $\Phi_{s-1}$.

If state $\Phi$ satisfies Fock's condition (\ref{condi}), 
so does $\Psi$ given by Eq. (\ref{cfwf}), and is therefore a legitimate state.
Thus composite fermion states with parallel or antiparallel flux attachment
can be constructed from electronic Hartree-Fock states as usual.
The CF theory for SU(2) systems thus generalizes trivially to SU($n$) systems.

\subsection{New FQHE states}

These observations enable us to construct 
variational wave functions for $\tilde\nu=\frac{m}{2pm\pm1}$.
For $m=1$, a single filled $\Lambda$ level (in one of the $\alpha^t$ spin bands) is the only candidate for an
incompressible ground state; the SU(4) multiplet is $[N]$.
For $m=2$, the two $\Lambda$ levels are filled with particles of identical 
or different spin state;
the multiplet is $[N]$ or $[N/2,N/2]$, respectively.
As the restriction on the orbital part of the CF wave functions is identical to the SU(2) case,
we expect the latter is prefered in the absence of a symmetry-breaking field.

\begin{figure}[!htbp]
\begin{center}
\includegraphics[width=0.85\columnwidth,keepaspectratio]{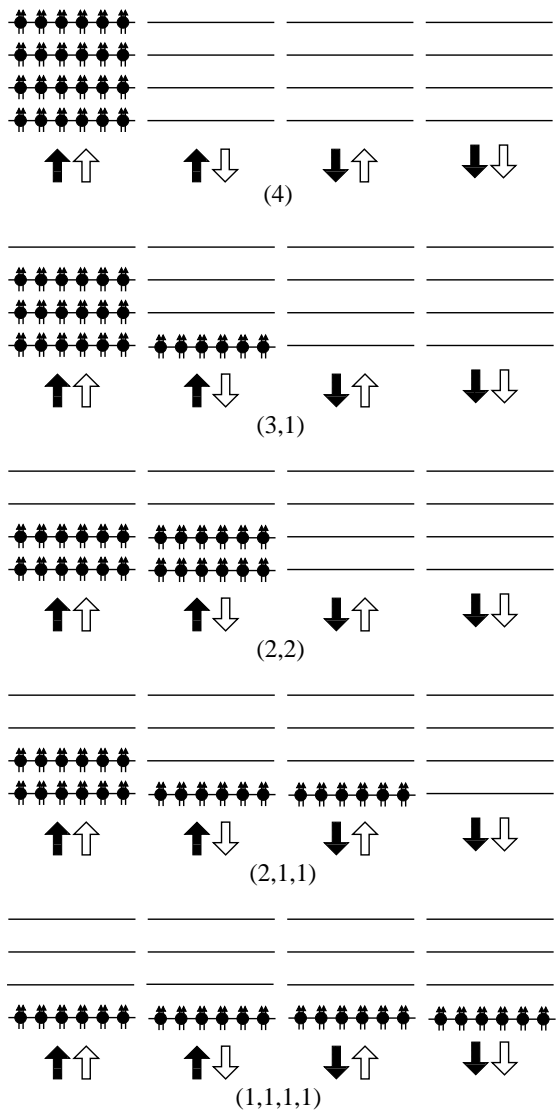}
\end{center}
\caption{\label{example}
The construction of composite fermion ground states for fractions where four $\Lambda$-levels can be filled
($\tilde\nu=\frac{4}{9}$, $\frac{4}{7}$, etc).
While the top three states are possible in SU(2) systems, the lowest two require an SU(4) symmetric internal degree
of freedom. The filled and empty arrows indicate the $z$-component of the spin $S_z$ and the pseudospin $P_z$, respectively.
This construction remains valid when $S'_z$ and $P'_z$ replace $S_z$ and $P_z$ (c.f.\ Eq.~(\ref{newspin})).  The top three states also occur in GaAs; the bottom three are 
new.
}
\end{figure}

For $m\ge3$, however, the SU(4) internal degree of freedom allows new ground state candidates.
Let $(l_1,l_2,l_3,l_4)$ denote a trial wave function that fills $l_1,\dots,l_4$ $\Lambda$ levels of the different single particle spin states.
As an example, Fig.~\ref{example} depicts all such states for $m=4$.
(For the monopole strength in finite spherical models, see Eq.~(\ref{mono}) below.)
Which of these states will be the ground state 
will be determined only by detailed microscopic calculations (to be described below).  A 
mean field approximation, which has been found to be successful for SU(2) electrons \cite{Wu93,Park1},
predicts that the state with the lowest ``CF kinetic energy" 
is the ground state.  This approximation, however, neglects the residual 
interaction between composite fermions, which favors the maximally polarized 
state due to exchange contribution.

\section{Exact diagonalization results}
\label{onethird}
\subsection{Ground states}

Table \ref{ground} shows the angular momentum and the SU(4) multiplet 
of the ground state in the $|n|=0,1$ Landau levels at flux values given by 
\begin{equation}
\label{mono}
2Q=2p(N-1)\pm\left(\frac{N-\sum_i l_i^2}{m}\right),
\end{equation}
where the $(l_1,l_2,l_3,l_4)$ state is a candidate for the ground state at $\tilde\nu=\frac{m}{2pm\pm1}$
with $m=\sum_i l_i$.
The following observations are consistent with the CF theory:

\begin{enumerate}

\item All ground states at the $N$ and $Q$ values satisfying Eq. (\ref{mono}) 
have rotational symmetry ($L=0$), consistent 
with the CF theory expectation of incompressibility for these parameters 
$N$ and $2Q$.  These systems therefore are finite size representations 
of incompressible states at $\tilde\nu=\frac{m}{2pm\pm1}$.

\item The filling $\tilde\nu=\frac{1}{3}$ corresponds to $m=1$, a single filled $\Lambda$ level
(in one of the $\alpha^t$ spin bands) is the only candidate for an
incompressible ground state; the SU(4) multiplet is $[N]$, which is consistent with Table \ref{ground}.
The energy is identical to the SU(2) case, 
but the multiplicity of the ground state is different ($\approx N^3/6$).
Analogous to $\tilde\nu=1$, the ground state is a completely symmetric $[N]$ multiplet,
which has associated with it 
a fully antisymmetric orbital wave function, in accordance with the Pauli's principle.
This is the SU(4) equivalent of the well-known ferromagnetic behavior in
systems with quadratic dispersion relation (GaAs heterostructures and quantum wells) in the vanishing Zeeman energy limit.
Unlike in the SU(2) symmetric case, however, the spins are not 
necessarily aligned in the same direction; 
in the $[N]$ multiplet spin and pseudospin are equal, but they can take all of the possible values\cite{Quesne,Hecht}:
$S=P=0,1,\dots,\frac{N}{2}$ if $N$ is even, and 
$S=P=\frac{1}{2},\frac{3}{2},\dots,\frac{N}{2}$ if $N$ is odd.
(The total multiplicity comes from these choices combined with the choice of $P_z$ and $S_z$.)
Any weak perturbation that polarizes the spin should also polarize the pseudospin.

\item
For $m=2$, two $\Lambda$ levels are filled with particles of identical or different spin state;
the multiplet is $[N]$ or $[N/2,N/2]$, respectively.
As the orbital part of the CF wave functions is identical to the SU(2) case,
it is no surprise that the latter is prefered in the absence of a symmetry-breaking field
(c.f.\ the $\nu=\frac{2}{3},\frac{2}{5}$ data in Table \ref{ground}).
(The multiplicity of $[\frac{N}{2},\frac{N}{2}]$ is $\approx N^4/96$.)
The ground state energy is the same as in the SU(2) symmetric case.
In Sec.~\ref{newstates} it was shown that the maximal weight state is a product of two Slater determinants
just as for an SU(2) singlet.
Because the SU(2) systems already find the lowest energy state in this class, 
the orbital part of the wave function remains the same.
However, the ground state is neither a spin nor a pseudospin singlet;
all spin $S$ and pseudospin $P$ combinations occur with $(S+P)=0,2,\dots,\frac{N}{2}$ for $N$ even, and
$(S+P)=1,3,\dots,\frac{N}{2}$ for $N$ odd.

\item At $\tilde\nu=\frac{3}{7}$ and $\tilde\nu=\frac{4}{9}$, the $n=0$ LL ground state multiplets are consistent with three and four
copies of completely filled lowest $\Lambda$ levels, respectively, consistent with 
the expectation from the mean-field model.
(Notice that $[N/3,N/3,N/3]$ and $[N/4,N/4,N/4,N/4]$ 
are conjugate representations to $[N/3]$ and $[0]$, respectively, which
cannot be distinguished by their $(S,P)$ quantum numbers.)
The mean-field expectation is not borne out in the $|n|=1$ LL.
In Sec.~\ref{newstates}, however, we have seen that new composite fermion ground states become possible at these fractions;
the small system calculations we can perform at
$\tilde\nu=\frac{3}{7}$ and $\tilde\nu=\frac{4}{9}$ 
are unable to determine the true nature of the ground states at these fractions.

\end{enumerate}

\begin{table}[htb]
\begin{center}
\begin{tabular}{r|r|r|r|c|c|c|c}
\hline\hline
$\tilde\nu$ & $N$ & $2Q$ & $D$ & $L^{(0)}$ & $SU(4)$ m. & $L^{(1)}$ & $SU(4)$ m. \\
\hline
$\frac{1}{3}$
      & 4  &  9 &    960 & 0 & [4] & 0 & [4] \\
      & 5  & 12 &  12934 & 0 & [5] & 0 & [5] \\
      & 6  & 15 & 282824 & 0 & [6] & 0 & [6] \\
\hline
$\frac{2}{3}$
      & 4  &  5 &    204 & 0 &[2,2]& 0 &[2,2]\\
      & 6  &  8 &  14464 & 0 &[3,3]& 0 &[3,3]\\
\hline
$\frac{2}{5}$
      & 4  &  7 &    488 & 0 &[2,2]& 0 &[2,2]\\
      & 6  & 12 &  97316 & 0 &[3,3]& 0 &[3,3]\\
\hline
$\frac{3}{7}$
      & 3  &  4 &     27 & 0 & [1] & 0 & [1] \\
      & 6  & 11 &  64392 & 0 & [2] & 0 & [4] \\
\hline
$\frac{4}{9}$
      & 4  &  6 &    325 & 0 & [0] & 0 & [4] \\
\hline\hline
\end{tabular}
\end{center}
\caption{\label{ground} 
Orbital angular momentum $L^{(n)}$ and $SU(4)$ multiplet of the ground states of finite systems on a sphere at the most prominent 
fractions in the $n=0$ and $|n|=1$ Landau levels of graphene.  The quantity $\tilde\nu$ 
is the filling factor within the Landau level; $Q$ is the monopole strength; $N$ is 
the number of electrons; and 
$D$ is the dimension of the Hilbert space in the $L_z=S_z=P_z=0$ sector.}
\end{table}

\subsection{Charged excitations: SU(4) CF skyrmions}

For SU(2) symmetry and vanishing Zeeman energy, the 
charged excitations at $\nu=1/3$, which map into filling factor one of composite 
fermions, are skyrmions of composite fermions \cite{CFSkyrmion}.   These are analogous to skyrmions of electrons \cite{Skyrmion} at $\nu=1$.
The skyrmions at these fillings are obtained as the ground states with the monopole strength $Q$ changed by $\pm1$
relative to the $Q$ of the sequence representing the state at $\tilde\nu$.
As Tables \ref{groundqp} and \ref{groundqh} testify, near $\tilde\nu=\frac{1}{3}$,
these states belong to an $\left[\frac{N}{2},\frac{N}{2}\right]$ or an $\left[\frac{N+1}{2},\frac{N-1}{2}\right]$
multiplet, depending on whether $N$ is even or odd.
(The degeneracies of these multiplets are $\approx N^4/96$ and $\approx N^4/48$, respectively.)
We identify these states with SU(4) composite fermion skyrmions.
The energies of these states are identical to the familiar SU(2) skyrmions,
 as these spinors enforce (cf.\ Sec.~\ref{newstates})
exactly the same partition of the wave function into antisymmetric factors as the SU(2) singlet ($N$ even) or spin-$\frac{1}{2}$ ($N$ odd).
Consequently, the gaps at $\tilde\nu=\frac{1}{2p+1}$ are identical to those calculated with SU(2) symmetry\cite{FQHEgraphene,CFSkyrmion}:
$\Delta_{1/3}^{(0)}=0.043(5)e^2/\epsilon l_B$ and $\Delta_{1/3}^{(1)}=0.017(3)e^2/\epsilon l_B$.

\begin{table}[htb]
\begin{center}
\begin{tabular}{r|r|r|r|c|c|c|c}
\hline\hline
$\tilde\nu$ & $N$ & $2Q$ & $D$ & $L^{(0)}$ & $SU(4)$ m. & $L^{(1)}$ & $SU(4)$ m. \\
\hline
1     & 4  &  2 &     25 & 0 &[2,2]& 0 &[0,0]\\
      & 5  &  3 &     92 & $\frac{1}{2}$ &[3,2]& $\frac{1}{2}$ &[3,2]\\
      & 6  &  4 &    644 & 0 &[3,3]& 0 &[3,3]\\
      & 7  &  5 &   3236 & $\frac{1}{2}$ &[4,3]& $\frac{1}{2}$ &[4,3]\\
      & 8  &  6 &  24483 & 0 &[4,4]& 0 &[4,4]\\
      & 9  &  7 & 142587 & $\frac{1}{2}$ &[5,4]& $\frac{1}{2}$ &[5,4]\\
\hline
$\frac{1}{3}$
      & 4  &  8 &    697 & 0 &[2,2]& 0 &[2,2]\\
      & 5  & 11 &   9296 & $\frac{1}{2}$ &[3,2]& $\frac{1}{2}$ &[3,2]\\
      & 6  & 14 & 203132 & 0 &[3,3]& 0 &[3,3]\\
\hline\hline
\end{tabular}
\end{center}
\caption{\label{groundqp} 
Orbital angular momentum $L^{(n)}$ and $SU(4)$ multiplet of the ground states of finite systems on the sphere
on the \emph{quasiparticle} side of $\nu=1$ and $\frac{1}{3}$ in the $n=0$ and $|n|=1$ Landau levels of graphene.
$D$ is the dimension of the Hilbert space in the $L_z=S_z=P_z=0$ sector.}
\end{table}

\begin{table}[htb]
\begin{center}
\begin{tabular}{r|r|r|r|c|c|c|c}
\hline\hline
$\tilde\nu$ & $N$ & $2Q$ & $D$ & $L^{(0)}$ & $SU(4)$ m. & $L^{(1)}$ & $SU(4)$ m. \\
\hline
1     & 4  &  4 &    117 & 0 &[2,2]& 0 &[2,2]\\
      & 5  &  5 &    521 & $\frac{1}{2}$ &[3,2]& $\frac{1}{2}$ & [3,2] \\
      & 6  &  6 &   3868 & 0 &[3,3]& 0 &[3,3]\\
      & 7  &  7 &  21170 & $\frac{1}{2}$ &[4,3]& $\frac{1}{2}$ & [4,3]\\
      & 8  &  8 & 165992 & 0 &[4,4]& 0 &[4,4]\\
\hline
$\frac{1}{3}$
      & 4  & 10 &   1281 & 0 &[2,2]& 0 &[2,2]\\
      & 5  & 13 &  17490 & $\frac{1}{2}$ &[3,2]& $\frac{1}{2}$ &[3,2]\\
      & 6  & 16 & 385784 &   &     &   &     \\
\hline\hline
\end{tabular}
\end{center}
\caption{\label{groundqh} 
Orbital angular momentum $L^{(n)}$ and $SU(4)$ multiplet of the ground states of finite systems on the sphere
on the \emph{quasihole} side of $\nu=1$ and $\frac{1}{3}$ in the $n=0$ and $|n|=1$ Landau levels of graphene.
$D$ is the dimension of the Hilbert space in the $L_z=S_z=P_z=0$ sector.}
\end{table}

As a technical remark, we note that the Hilbert space in the calculations leading to the 
result of Tables \ref{ground}-\ref{groundqh} can be further reduced 
by exploiting the existence of a third SU(4) generator, traditionally called $E_{00}$, that commutes with the Hamiltonian.
We identify SU(4) multiplets only by their $(S,P)$ quantum numbers\cite{Hecht,Quesne},
and do not implement this simplification.

\section{Thermodynamic limit}
\label{thermo}

Because of the four-fold degeneracy of the graphene Landau levels, exact 
diagonalization studies are possible only for very small systems.  Having established the 
regime of validity of the CF theory, we now use the wave functions of Eq.~(\ref{cfwf}) 
to study large systems by the Monte Carlo method, and obtain 
thermodynamic limits for the energies of various candidate states.

For the $n=0$ Landau level the computation is straightforward.  
In the $|n|=1$ Landau levels the calculation of the energy expectation value by the Monte Carlo method 
requires a knowledge of the real-space interaction corresponding to the effective pseudopotentials $V^{(1)\textrm{gr.}}_m$.
Following a well-tested procedure\cite{secondll,Park98}, we use a convenient form,
\begin{equation}
V^{\text{eff}}(r)=\frac{1}{r}+\sum_{i=0}^M c_i r^i e^{-r},
\label{form}
\end{equation}
and fit the coefficients $c_i$ to reproduce the first $M+1$ pseudopotentials $V^{(1)\textrm{gr.}}_m$ of Eq.~(\ref{effective}) by
\begin{equation}
V^{(n)\text{eff}}_m=\frac{\langle\eta_{0,m}|\sum_{i<j}V^{\text{eff}}(z_i-z_j)|\eta_{0,m}\rangle}{\langle\eta_{0,m}|\eta_{0,m}\rangle}.
\end{equation}
The coefficients $c_i$ thus obtained are given in Table \ref{coeff}.
Our effective interaction produces the $V_0^{(n)\textrm{gr.}},\dots,V_6^{(n)\textrm{gr.}}$ pseudopotentials exactly; we 
have checked that the relative error in the remaining pseudopotentials is very small
and does not affect our conclusions.

\begin{table}
\caption{\label{coeff} Coefficients in Eq.~(\ref{form}), which reproduces the effective interaction for the $|n|=1$ LL of graphene.}
\begin{ruledtabular}
\begin{tabular}{rl}
Coefficient & Value \\ \hline
$c_0$ & -11.1534 \\
$c_1$ & 24.9078 \\
$c_2$ & -18.6461 \\
$c_3$ & 6.63657 \\
$c_4$ & -1.221097 \\
$c_5$ & 0.112068 \\
$c_6$ & -0.00404269
\end{tabular}
\end{ruledtabular}
\end{table}

\begin{figure*}[!htbp]
\begin{center}
\includegraphics[width=\textwidth,keepaspectratio]{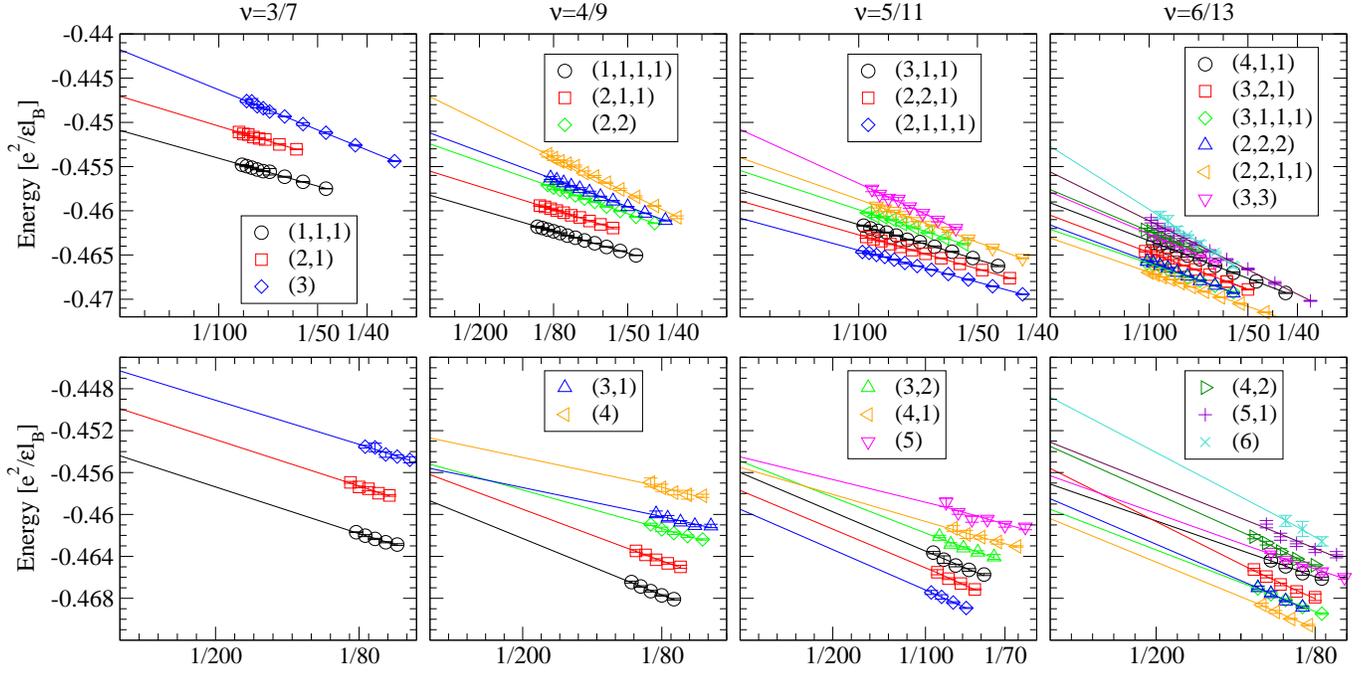}
\end{center}
\caption{\label{groundp1} [Color online]
Energy per particle for the novel incompressible composite fermion ground state
wave functions for the sequence $\tilde\nu=m/(2pm+1)$.  The states are denoted by 
$(l_1,l_2,l_3,l_4)$, explained in the text.  The upper panels are 
for the $n=0$ Landau level, and the lower panels for the $|n|=1$ Landau level.  The 
energies are given in units of $e^2/\epsilon l_B$, where $l_B=\sqrt{\hbar c/eB}$ is the 
magnetic length.  The thermodynamic limits of the 
the ground state energies are given in Table \ref{groundtab}.}
\end{figure*}

We have studied composite fermion wave functions for $\tilde\nu=\frac{m}{2pm\pm1}$.
 Figure \ref{groundp1} shows the energies as a function of $N$ for all ground state candidates at these fractions, and the 
thermodynamic limits of the energies are given in Table \ref{groundtab}.  

For the sequences $\tilde\nu=\frac{m}{2m+1}$, which represent the integral 
quantum Hall effect of $^2$CFs (composite fermions carrying two vortices),
the ground state is consistent with the prediction of the mean-field approximation:
it is the one that best exploits the SU(4) spin bands to minimize the CF kinetic 
energy.  The examples are:
the state $(1,1,1)$ at $\tilde\nu=\frac{3}{7}$; the SU(4) singlet state $(1,1,1,1)$ at $\tilde\nu=\frac{4}{9}$; 
state $(2,1,1,1)$ at $\tilde\nu=\frac{5}{11}$; and 
state $(2,2,1,1)$ at $\tilde\nu=\frac{6}{13}$.

\begin{figure*}[!htbp]
\begin{center}
\includegraphics[width=\textwidth,keepaspectratio]{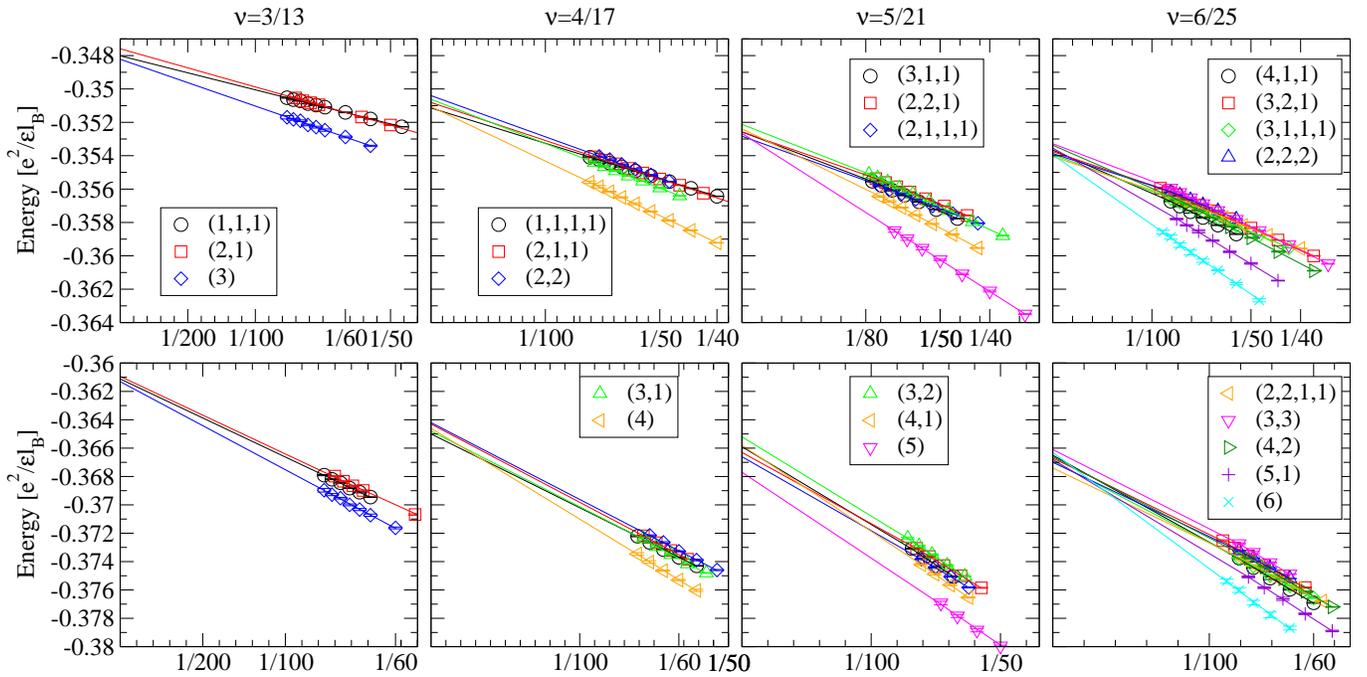}
\end{center}
\caption{\label{groundp2} [Color online]  Same as in Fig. \ref{groundp1}, but 
for the sequence $\tilde\nu=m/(4m+1)$, corresponding to the integral 
quantum Hall effect of $^4$CFs.
The thermodynamic limits of the 
the ground state energies are given in Table \ref{groundtab2}.}
\end{figure*}

For the sequence $\tilde\nu=\frac{m}{4m+1}$, appropriate for $^4$CFs, 
the mean-field approximation is not always valid.  The 
ferromagnetic CF ground states are competitive even when they do not 
have the lowest CF kinetic energy, as seen in \ Figure \ref{groundp2} and 
Table \ref{groundtab2}.  The ground state
at $\tilde\nu=\frac{3}{13}$ is found to be fully
polarized; the state with the least possible CF kinetic
energy per particle has slightly higher energy.
At $\tilde\nu=\frac{4}{17}$, $\frac{5}{21}$ and $\frac{6}{25}$ the states $(1,1,1,1)$,
$(2,1,1,1)$ and $(2,2,1,1)$, respectively, win by a very small margin.
Apparently, the gain from CF kinetic energy minimization is on the 
same order as that from exchange maximization; which state becomes the 
ground state is determined by their competition.  Our results demonstrate 
that the inter-CF interaction are stronger for $^4$CFs than for $^2$CFs.

For the SU(2) case, it had been found \cite{Park2} that for  
$\nu=\frac{m}{2m+1}$ the model of  non-interacting composite fermions 
predicts the ground state quantum numbers correctly, which are also 
confirmed extensively through several experiments.  For $\nu=\frac{m}{4m+1}$, on 
the other hand, a fully spin polarized state was found to have the lowest energy 
in detailed calculations with the CF theory\cite{Park2}.
The behavior is more complicated for the SU(4) symmetry, where, at least for 
some fractions, the fully polarized states are not the ground states.

The data in Tables \ref{groundtab} and \ref{groundtab2}
include the ground states that are available in SU(2) symmetric systems.
Our results are consistent with those of Park and Jain \cite{Park1} for the SU(2) case,
although the values differ slightly; the current results are more accurate.
The extrapolation to the theormodynamic limit in our work 
is based on $N=36-104$ particles,
with linear functions fitted on the $E(1/N)$ curve
under the condition $\chi^2\lesssim N-2$.
 
It is in principle possible to apply similar methods to 
the states at $\tilde\nu=\frac{m}{2pm-1}$, where the effective magnetic field $B^\ast$ felt by composite fermions is antiparallel to the real external field $B$.
The projection procedure for CF wave functions has been elaborated by M\"oller and Simon \cite{Moller}, but its implementation is impractical  
for the number of $\Lambda$ levels and effective monopole strengths $q$ that are of interest.  (The maximal degree of derivatives for negative-$B^*$ states is
\cite{Moller} $2|q|+m$, while for parallel flux attachment it is only $m$.)
Therefore, we do not pursue that direction in the present work.

\begin{table}[htb]
\begin{center}
\begin{tabular}{c|c|l|l|c}
\hline\hline
$\tilde\nu$ & State & Energy $n=0$ & Energy $n=1$ & $\hbar\omega_c^\ast/N$\\
\hline
$\frac{3}{7}$
& (3)     & -0.44178(8)& -0.4463(14)& 1 \\
& (2,1)   & {-0.44704(15)}& {-0.4499(7)} & 1/3 \\
& (1,1,1) & {-0.45090(3)} & {-0.4544(4)} & 0 \\
\hline
$\frac{4}{9}$
& (4)       & -0.4471(2) & -0.4527(15) & 3/2 \\
& (3,1)     & {-0.4512(1)}  & {-0.4556(12)} & 3/4 \\
& (2,2)     & {-0.45244(9)} & {-0.4552(6)}  & 1/2 \\
& (2,1,1)   & {-0.45552(12)}& {-0.4562(7)}  & 1/4 \\
& (1,1,1,1) & {-0.45825(5)} & {-0.4587(5)}  & 0 \\
\hline
$\frac{5}{11}$
& (5)       & -0.4508(4)  & -0.4545(9)  & 2 \\
& (4,1)     & {-0.45399(11)}& {-0.4555(7)} & 6/5 \\
& (3,2)     & {-0.45544(16)}& {-0.4549(2)} & 4/5 \\
& (3,1,1)   & {-0.45762(8)} & {-0.4560(6)} & 3/5 \\
& (2,2,1)   & {-0.45888(7)} & {-0.4577(8)} & 2/5 \\
& (2,1,1,1) & {-0.46084(5)} & {-0.4595(9)} & 1/5 \\
\hline
$\frac{6}{13}$
& (6)       & -0.4527(4) & -0.4488(19) & 5/2 \\
& (5,1)     & {-0.4556(1)}  & {-0.4531(14)} & 5/3 \\
& (4,2)     & {-0.4576(4)}  & {-0.4535(7)}  & 7/6 \\
& (3,3)     & {-0.4579(2)}  & {-0.4563(6)}  & 1 \\
& (4,1,1)   & {-0.4591(1)}  & {-0.4571(13)} & 1 \\
& (3,2,1)   & {-0.4605(1)}  & {-0.4556(8)}  & 2/3 \\
& (2,2,2)   & {-0.46161(9)} & {-0.4585(9)}  & 1/2 \\
& (3,1,1,1) & {-0.4621(1)}  & {-0.4595(6)}  & 1/2 \\
& (2,2,1,1) & {-0.46308(4)} & {-0.4604(7)}  & 1/3 \\
\hline\hline
\end{tabular}
\end{center}
\caption{\label{groundtab} 
Energy per particle for incompressible composite fermion ground state candidates of different SU(4) symmetries at $\tilde\nu=\frac{m}{2m+1}$,
corresponding to the integral quantum Hall effect of $^2$CFs.
For each fraction $\tilde\nu$ the states are listed in decreasing order of energy.
The last column gives the CF kinetic energy ($\hbar \omega_c^*$) 
per particle in the thermodynamic limit, measured relative to the energy of the 
lowest $\Lambda$ level.
}
\end{table}

\begin{table}[htb]
\begin{center}
\begin{tabular}{c|c|l|l|c}
\hline\hline
$\tilde\nu$ & State & Energy $n=0$ & Energy $n=1$ & $\hbar\omega_c^\ast/N$ \\
\hline
$\frac{3}{13}$
& (2,1)   & {-0.34759(4)}  & {-0.36094(17)} & 1/3 \\
& (1,1,1) & {-0.34801(3)}  & {-0.36112(17)} & 0 \\
& (3)     & -0.34822(12) & -0.36131(29) & 1 \\
\hline
$\frac{4}{17}$
& (2,2)     & {-0.35040(8)} & {-0.3642(2)} & 1/2 \\
& (3,1)     & {-0.35062(7)} & {-0.3648(2)} & 3/4 \\
& (2,1,1)   & {-0.35081(3)} & {-0.3643(2)} & 1/4 \\
& (4)       & -0.35106(5) & -0.3646(3) & 3/2 \\
& (1,1,1,1) & {-0.35114(3)} & {-0.3650(1)} & 0 \\
\hline
$\frac{5}{21}$
& (3,2)     & {-0.35212(5)} & {-0.3652(3)} & 4/5 \\
& (4,1)     & {-0.35238(9)} & {-0.3658(4)} & 6/5 \\
& (3,1,1)   & {-0.35257(5)} & {-0.3659(3)} & 3/5 \\
& (2,2,1)   & {-0.35257(5)} & {-0.3663(1)} & 2/5 \\
& (5)       & -0.35266(10)& -0.3677(3) & 2 \\
& (2,1,1,1) & {-0.35283(6)} & {-0.3666(3)} & 1/5 \\
\hline
$\frac{6}{25}$
& (3,3)     & {-0.35328(2)} & {-0.3661(2)} & 1 \\
& (4,2)     & {-0.35337(6)} & {-0.3666(2)} & 7/6 \\
& (5,1)     & {-0.35354(9)} & {-0.3668(4)} & 5/3 \\
& (4,1,1)   & {-0.35366(5)} & {-0.3665(1)} & 1 \\
& (3,2,1)   & {-0.35375(3)} & {-0.3667(1)} & 2/3 \\
& (2,2,2)   & {-0.35387(7)} & {-0.3670(2)} & 1/2 \\
& (3,1,1,1) & {-0.35393(3)} & {-0.3669(1)} & 1/2 \\
& (6)       & -0.35394(18)& -0.3663(6) & 5/2 \\
& (2,2,1,1) & {-0.35405(4)} & {-0.3674(2)} & 1/3 \\
\hline\hline
\end{tabular}
\end{center}
\caption{\label{groundtab2} Same as in Table \ref{groundtab} for $\tilde\nu=\frac{m}{4m+1}$,
corresponding to the integral quantum Hall effect of $^4$CFs.
Energies per particle for incompressible composite fermion ground state candidates with different SU(4) symmetries are shown in decreasing order.
For each fraction $\tilde\nu$ the states are in decreasing order of energy.
}
\end{table}

\section{Quantum phase transitions}
\label{phasetrans}

Either the Zeeman energy $E_Z^S=q\mu_B\vec B\cdot\vec S$, or the pseudo-Zeeman energy $E_Z^P=\Delta P_z$, or both,  
will break the SU(4) symmetry. Assuming the symmetry-breaking fields are weak,
the effect will be to select the most favorable member of the ground state multiplet.
(Notice that $P_z$, $S_z$, as well as the third member $E_{00}$ of the Abelian subalgebra
of SU(4) generators, commute with the Hamiltonian of Eq.~(\ref{hamiltonian}) in the SU(4)
symmetric $\Delta\to0, g\to0$ limit.)
Slightly stronger fields may drive zero-temperature phase transitions between
the possible CF ground states at a fixed filling factor.
It is convenient to change to new quantum numbers $S'_z,P'_z$ by
\begin{equation}
\label{newspin}
\begin{pmatrix}
S'_z \\ P'_z
\end{pmatrix}=U
\begin{pmatrix}
S_z \\ P_z
\end{pmatrix},
\end{equation}
with a unitary $U$ to eliminate one kind of Zeeman energy.
Choosing
\begin{equation}
U=\frac{1}{\sqrt{1+\lambda^2}}
\begin{pmatrix}
1 & \lambda \\
-\lambda & 1
\end{pmatrix},
\end{equation}
with $\lambda=E^P_Z/E^S_Z$ yields
\begin{eqnarray*}
{E_Z^S}'&=&\frac{1}{\sqrt{1+\lambda^2}}E^S_Z + \frac{\lambda}{\sqrt{1+\lambda^2}}E^P_Z,\\
{E_Z^P}'&=&0.
\end{eqnarray*}
The phase transitions driven by the effective ${E_Z^S}'$ are given in Table \ref{transitab}.
Here we assume a sufficiently weak field that does not mix the low-lying
states, but only selects the most favorable member of the ground state multiplet.
In state $(l_1,l_2,l_3,l_4)$ the with $l_i\ge l_{i+1}$, this means
filling $l_1+l_2$ $\Lambda$ levels of the favorable $S'$ spin, and $l_3+l_4$ levels of the
unfavorable $S'$ spin.
This results in an effective Zeeman energy difference per particle.
{Notice that no value of $E^P_Z$ and $E^S_Z$ will drive the system at $\tilde\nu=m/(2pm+1)$ to a completely antisymmetric orbital state $(m)$.
This is a consequence of our freedom to choose a basis $P'_z,S'_z$ in the Cartan subalgebra of SU(4) (Eq.~(\ref{newspin})),
which eliminates one kind of Zeeman energy; the states where the two unfavorable bands of the $S'$ spin are emptied
will already have the lowest possible effective Zeeman energy ${E_Z^S}'$.}

Tilting the magnetic field is often used as a means to tune the effective Zeeman energy.
Unlike in GaAs/AlGaAs heterostructures, the in-plane megnetic field is unlikely to change
the single-particle states in any significant manner, because the transverse thickness
of the two dimensional electron system is graphene ($\sim 2$\AA) is much smaller than the typical magnetic lengths ($\sim$100\AA).
The most favorable parameter space for 
the observation of these transitions occurs for very low values of 
the dielectric constant $\epsilon$ and the sublattice asymmetry $\Delta$.
The value of $\epsilon$ depends on the interaction with the substrate
on which the graphene sheet is placed.

\begin{table}[htb]
\begin{center}
\begin{tabular}{c|l|l|l}
\hline\hline
$\tilde\nu$ & Transition & $\left(\frac{{E_Z^S}'}{e^2/\epsilon l_B}\right)_c$ & Change of $\langle S'\rangle$\\
\hline
$\frac{3}{7}$  & $(1,1,1)\to(2,1)$     & {0.0116(6)} & $\frac{1}{6}\to\frac{1}{2}$ \\
\hline
$\frac{4}{9}$  & $(1,1,1,1)\to(2,1,1)$ & {0.0109(7)} & $0\to\frac{1}{4}$ \\
               & $(2,1,1)\to(2,2)$     & {0.0123(8)} & $\frac{1}{4}\to\frac{1}{2}$ \\
\hline
$\frac{5}{11}$ & $(2,1,1,1)\to(2,2,1)$ & {0.0098(6)} & $\frac{1}{10}\to\frac{3}{10}$ \\
               & $(2,2,1)\to(3,2)$     & {0.0172(8)} & $\frac{3}{10}\to\frac{1}{2}$ \\
\hline
$\frac{6}{13}$ & $(2,2,1,1)\to(3,2,1)$ & {0.0076(5)} & $\frac{1}{6}\to\frac{1}{3}$ \\
               & $(3,2,1)\to(3,3)$     & {0.016(2)}  & $\frac{1}{3}\to\frac{1}{2}$ \\
\hline\hline
\end{tabular}
\end{center}
\caption{\label{transitab}
Zero temperature phase transitions driven by the effective Zeeman energy in the $n=0$ Landau level.
The last column gives the expectation value of the transformed spin $S'$ (Eq.~(\ref{newspin})),
which coincides with the ordinary spin in the case of perfect sublattice symmetry ($\Delta=0$).
The quantity $\left({{E_Z^S}'}/{(e^2/\epsilon l_B)}\right)_c$ give the 
value of the parameter where the transition is predicted to take place.
}
\end{table}

\section{Conclusion}
\label{conclu}

We have used a combination of exact diagonalization and the CF theory to  identify a large range of 
possible FQHE states in graphene allowed by SU(4) symmetry,  and shown that  
new states, which have no analog in GaAs, can occur 
at filling factors $\tilde\nu=\frac{m}{2pm\pm1}$ for $m\ge 3$ in the $|n|=0$ and 
$|n|=1$ Landau levels of graphene.  For $^2$CFs, the ground states are those for 
which the composite fermion kinetic energy is minimum; these states 
spread out maximally in SU(4) spin space.
For $^4$CFs the fully polarized ground state wins, with the exception of 
$\tilde\nu=\frac{4}{17}$,
where very compact SU(4) singlet structures are possible.
We have also estimated parameter regimes where these states should occur;
zero temperature phase transitions can be driven by variation of 
an effective Zeeman energy that accounts for the combined effect of the Zeeman
 energy and the sublattice symmetry breaking field.
At $\tilde\nu=\frac{1}{3}$ we found SU(4) skyrmions to be the lowest 
energy charged excitations.

We thank the High Performance Computing (HPC) group at Penn State University ASET (Academic Services and Emerging Technologies)
for assistance and computing time on the Lion-XO cluster.
Partial  support of this research by the National Science Foundation under
grant No.\ DMR-0240458 is gratefully acknowledged.

\newcommand{\PRL}{Phys.\ Rev.\ Lett.}
\newcommand{\PRB}{Phys.\ Rev.\ B}
\newcommand{\NPB}{Nucl.\ Phys.\ B}

\end{document}